\documentclass[11pt]{article}
\pdfoutput=1
\usepackage{amssymb,amsmath,mathrsfs,enumerate}
\usepackage{graphicx,rotate,multicol}
\usepackage{caption,verbatim}
\usepackage{cite}
\usepackage[colorlinks=true,
linkcolor=red,
urlcolor=blue,
citecolor=blue]{hyperref}
\usepackage{color}
\usepackage{textcomp} 
\usepackage{setspace}
\usepackage{subfig}
\usepackage{verbatim}
\usepackage{wrapfig}
 
\evensidemargin=\oddsidemargin
\allowdisplaybreaks

\definecolor{darkgreen}{cmyk}{1,0,1,0.4}

\def\gmin2{$(g-2)_\mu$}

\begin{document}
	
	\begin{flushright}
	CUMQ/HEP 205 \\
	IPMU-20-0088 
	\end{flushright}

	\begin{center}
		{\Large \bf 
		  Muon Anomalous Magnetic Moment in Two Higgs Doublet Models with Vector-Like Leptons 	} \\
		\vspace*{1cm}  
		~Mariana Frank$^{a,}$\footnote{mariana.frank@concordia.ca},
		~Ipsita Saha$^{b,}$\footnote{ipsita.saha@ipmu.jp}\\
		\vspace{10pt} {\small } {\em 
			$^a$ Department of Physics,  
			Concordia University, 7141 Sherbrooke St. West,\\
			Montreal, Quebec, Canada H4B 1R6. \\
			$^b$ Kavli IPMU (WPI), UTIAS, University of Tokyo, Kashiwa, Chiba 277-8583, Japan. }   
	\end{center}
	\normalsize
	\begin{abstract}
		We show that  inclusion of a single generation of vector-like leptons in the
		Two-Higgs Doublet Models significantly enlarges the allowed parameter space consistent with the muon anomalous magnetic moment,  
		as well as with other theoretical and experimental constraints. While previously $(g -2)_\mu$ could only be resolved in Type-X scenario by requiring a light
		pseudoscalar Higgs boson and large $\tan \beta$, contributions of 
	   vector-like leptons via  two-loop Barr Zee diagrams  
	    broaden the allowed parameter space, allowing $\tan \beta$ as low as 10 and pseudoscalar masses as large as ${\cal O} (1~\rm TeV)$, while fulfilling 
	    the stringent constraints from  precision and flavor observable. Similar results are obtained for Type-II scenarios, but there the parameter space is more restricted by flavor observables.
	\end{abstract}
	\bigskip
	\section{Introduction}
	\label{sec:intro}
	The anomalous magnetic moment of the muon, representing the deviation from $g_\mu=2$, is one of the most debated topics in the
	field of particle phenomenology. The magnetic moment of the muon is among the most 
	accurately predicted quantities within the Standard Model (SM), and is very precisely measured. 
Comparison between experiment and theory tests the
SM at loop levels, with any deviation from the SM expectation interpreted as a signal of new physics \cite{Czarnecki:2001pv},  with current sensitivity reaching up to mass scales of ${\cal O}$(TeV) \cite{Jegerlehner:2009ry,Miller:2007kk,Miller:2012opa}.  There is a long-standing discrepancy between  the theoretical prediction and the measured value of the anomalous magnetic moment of the muon \cite{Bennett:2006fi}	
	\begin{equation}
	a_\mu^{\rm exp}=116592091(54)(33) \times 10^{-11}\, ,
	\end{equation}
where the errors in brackets are systematic, and then statistical. The latest world average of the predicted value from the Standard Model (SM) is given by \cite{Aoyama:2020ynm,Davier:2019can}
	\begin{equation}
	a_\mu^{\rm SM} = 11 659 1810(43) \times 10^{-11}~,
	\label{gmt-sm}
	\end{equation}	
	where the errors are from electroweak, lowest-order hadronic, and higher-order hadronic contributions.
	The difference 
	\begin{eqnarray}
 \Delta a_\mu= a_\mu^{\rm exp}- a_\mu^{\rm SM}=(281 \pm 76) \times 10^{-11}\,,
 \label{eq:gmin2}
	\end{eqnarray}
     indicates a 3.7$\sigma$ discrepancy between theory and experiment, and is interpreted as an indication of new physics. This discrepancy will be further explored at Fermilab \cite{Grange:2015fou} and J-PARC \cite{Iinuma:2011zz} experiments in the near future. 
	
	The possible reason for the current deviation  between the SM and the experimental value has been explored in numerous new physics models in the past few decades. In particular,  extending the fermion sector of the SM by the addition of a vector-like lepton (VLL) generation can easily explain the $(g-2)_\mu$ discrepancy~\cite{Dermisek:2013gta,Falkowski:2013jya}.  However, in this scenario muon mixing with the new VLLs is  necessary, and this alters significantly  the Higgs decay branching ratio to muons, as well as the Higgs to diphoton decay branching ratio (BR), violating the constraints imposed by the current collider Higgs data~\cite{Dermisek:2013gta}. Moreover, additional constraints are also put forward by flavor observable~\cite{Ishiwata:2015cga}. Thus to be able to consistently explain the $(g-2)_\mu$ discrepancy with VLLs, one must  resort to models beyond the SM.
	
As minimal scalar sector extensions of the SM,
	the Two-Higgs Doublet Models (2HDM) ~\cite{Branco:2011iw,Bhattacharyya:2015nca,Cao:2009as}, are of particular interest. 
	In these models, there are two neutral CP-even Higgs bosons (the SM $h$ and heavier $H$), one pseudoscalar state ($A$) and one charged Higgs boson ($H^\pm$). A $Z_2$ symmetry is imposed to avoid flavor changing neutral current (FCNC)  interactions at tree-level~\cite{Glashow:1976nt}, and then the models are classified according to the $Z_2$ charge assignments for the SM fermions. Among the four variants, only the Type-X and Type-II models are able to explain the anomalous nature of the $(g-2)_\mu$. Due to larger couplings of leptons to the additional non-SM Higgs bosons, these 2HDM models can solve the anomaly by including, in addition to the usual one-loop contributions,  two-loop contributions from the Barr Zee type diagrams \cite{Broggio:2014mna,Lindner:2016bgg,Wang:2018hnw}.  These diagrams contribute with the same order of magnitude, and sometimes dominate  the one-loop contribution, especially the contribution coming from the heavier Higgs bosons, which is large \cite{Ilisie:2015tra}.
	 The main difference between the Type-II and Type-X 2HDMs lies in the quark
     couplings to the additional Higgs bosons.  In Type-II 2HDM, both charged lepton and down-type quark couplings are proportional to  $\tan\beta$, and thus the model is severely constrained by flavor physics \cite{Lindner:2016bgg} and direct searches of extra Higgs bosons. The solution for the muon $g-2$ requires large values of $\tan \beta$ and very light pseudoscalar Higgs boson masses, $m_A$ which, in Type -II, are disallowed  from $B$-physics observables~\cite{Chun:2015xfx}.  By contrast, in Type-X 2HDM, the quark couplings to the extra  Higgs bosons are suppressed, while the lepton couplings are enhanced, and  
     the flavor constraints are weaker than in Type-II 2HDM. Thus, without additional fermion content, only  Type-X 2HDM survives as a possible framework for providing a solution to the anomalous magnetic moment of the muon, while maintaining consistency with the flavor constraints. Still, even in Type-X, agreement with low-energy data requires  very light $m_A$ and large $\tan \beta$ \cite{Broggio:2014mna,Wang:2014sda,Chun:2015xfx,Abe:2015oca,Chun:2015hsa,Wang:2018hnw}. Furthermore, with large $\tan \beta$, the Type-X model 
     becomes leptophilic and is stringently constrained by lepton precision observables ~\cite{Chun:2016hzs}.

	To remedy the above-mentioned shortcomings, we introduce a single generation of vector-like leptons (VLLs) into the 2HDM scenario \footnote{Such VLL extensions but either in the context of an extra inert doublet~\cite{Barman:2018jhz,Barman:2019oda,deJesus:2020upp,Jana:2020joi,Chen:2020tfr} or in generalized two-Higgs doublet scenario~\cite{Jana:2020pxx}, 2HDM with extra singlet~\cite{Sabatta:2019nfg} have been studied in the recent past.}. We show that this ameliorates both the above-mentioned problems of the individual models while still explaining the \gmin2 in a minimal approach. In our scenario, VLLs  couple with all the Higgs bosons, and their mass is generated by both the Yukawa and bare mass term.  In this model, mixing between these additional VLLs and the SM leptons is not necessary, and thus the Higgs signal constraints on dimuon decays will not be affected. Furthermore, we shall show that the SM Higgs decay to $\gamma \gamma$ will remain well within the experimental uncertainty by the contributions from the VLL loops and the charged Higgs scalar loop. Finally, the VLL coupling with the non-SM Higgs bosons gives rise to additional Barr Zee contributions, enhancing the value for \gmin2 even for a heavier pseudoscalar Higgs boson. This scenario can thus be viewed as a minimal set-up where VLLs resolve the \gmin2  anomaly, while maintaining consistency with theoretical and experimental constraints.
	
	Our paper is organized as  follows. In Sec.~\ref{sec:model}, we discuss the model Lagrangian for a $Z_2$ symmetry 2HDM potential augmented by a VLL generation. Following this, in Sec.~\ref{sec:constraints}, we explore all the relevant experimental and theoretical constraints including a 
	detailed analysis on the oblique parameter constraints and the Higgs to diphoton decay mode. Next, in Sec.~\ref{sec:g-2}, we address the contributions of all the relevant one-loop and two-loop Barr Zee diagrams to the muon anomalous magnetic moment in this   scenario. In Sec.~\ref{sec:results}, we discuss our results and findings, and address the effects of VLLs on heavy Higgs searches. Finally, we summarize and conclude  in Sec.~{\ref{sec:summary}.

\section{Two-Higgs Doublet Models with VLLs}
	\label{sec:model}
	
The scalar sector of the 2HDM is composed of two $SU(2)_L$ doublet scalar fields $H_1$ and $H_2$. 
To avoid  tree level FCNC, we introduce an extra $Z_2$ symmetry
under which the fields transform as $H_1 \to H_1$ and $H_2 \to - H_2$.  Labelling the $H_2$ as the Higgs field which couples to the up-type quark, the choices of $Z_2$ charge assignments for  different
fermion fields  lead to four different variants of the 2HDMs,
namely, Type-I, II, X and Y~\cite{Branco:2011iw}. We  focus on the Type-X set-up of the Yukawa interaction, as being most promising for \gmin2, but also investigate the consequences of our analysis on the Type-II structure in the respective section. 
In Type-X 2HDM, both the up and down-type quarks couple with $H_2$ while only the charged lepton couples with $H_1$.

The most general Higgs potential in 2HDM with  softly-broken $Z_2$ parity is 
\begin{eqnarray}
V (H_1, H_2) &= &
m_{11}^2 H_1^\dagger H_1
+ m_{22}^2 H_2^\dagger H_2
- ( m_{12}^2 H_1^\dagger H_2 + h.c.) 
+ \frac{\lambda_1}{2} \left(H_1^\dagger H_1\right)^2 
 + \frac{\lambda_2}{2} \left(H_2^\dagger H_2\right)^2 \nonumber\\
&+ &\lambda_3  \left(H_1^\dagger H_1\right)\left(H_2^\dagger H_2\right)
+ \lambda_4 \left(H_1^\dagger H_2\right)\left(H_2^\dagger H_1\right)
+ \left[ \frac{\lambda_5}{2} (H_1^\dagger H_2)^2 + h.c. \right]\,.
\label{eq:higgspotential}
\end{eqnarray}
Here, we consider  CP conservation in the scalar sector and 
assume all the parameters to be real.
The Higgs doublets are, in terms of  their component fields
\begin{eqnarray}
H_i = \left(\begin{array}{c}
h_i^+ \\
\frac{1}{\sqrt{2}}\left(v_i + h_i + i a_i\right) \\ 
\end{array}\right) \, ,\qquad (i=1,2) \,.
\label{eq:higgsfield}
\end{eqnarray}
The minimization condition for the potential $V (H_1, H_2)$ can be used to express the two bilinear terms in the potential  as functions of the two vacuum expectation values (VEVs) $v_1$ and $v_2$, where  we define $\tan\beta = v_2/v_1$ as their ratio.
The mass eigenstates of the scalar bosons are expressed by introducing the mixing angles $\alpha$ and $\beta$ as
\begin{eqnarray}
\left(\begin{array}{c}
h_1 \\ h_2
\end{array}\right) 
&=&
\left(\begin{array}{cc}
\cos\alpha & -\sin\alpha \\
\sin\alpha & \cos\alpha
\end{array}\right)
\left(\begin{array}{c}
H \\ h
\end{array}\right),\qquad
\left(\begin{array}{c}
a_1 \\ a_2
\end{array}\right) 
=
\left(\begin{array}{cc}
\cos\beta & -\sin\beta \\
\sin\beta & \cos\beta
\end{array}\right)
\left(\begin{array}{c}
G^0 \\ A
\end{array}\right),\\
\left(\begin{array}{c}
h_1^\pm \\ h_2^\pm
\end{array}\right) 
&=&
\left(\begin{array}{cc}
\cos\beta & -\sin\beta \\
\sin\beta & \cos\beta
\end{array}\right)
\left(\begin{array}{c}
G^\pm \\ H^\pm
\end{array}\right), 
\label{eq:higgseigenstates}
\end{eqnarray}
where $G^0$ and $G^\pm$ are the Nambu-Goldstone bosons and ($h\,,H\,,A\,,H^\pm$) are the two CP-even, one CP-odd
and the charged 
Higgs mass eigenstates, respectively, and where we assume $m_h < m_H$\footnote{For detailed mass and coupling relations, 
	refer to ~\cite{Branco:2011iw,Das:2015mwa}.}.

The general Yukawa interaction Lagrangian for 2HDM with SM fermions is given by 
\begin{equation}
{\cal L}_{\rm Y} = -y_u \tilde H_u^T Q_L u_R^c - y_d H_d^\dagger Q_L d_R^c - y_\ell H_\ell^\dagger \psi_L e_R^c + h.c.,\label{eq:yukawa}
\end{equation}
where $\tilde H_u = i\tau^2 H_u$.
In Eq.~(\ref{eq:yukawa}), $H_u$, $H_d$ and $H_\ell$ are either $H_1$ or $H_2$ depending on the variant of 2HDM chosen. 
As mentioned, for Type-X scenario, $H_u,H_d \equiv H_2$ while $H_\ell \equiv H_1$.
Working in the scalar bosons mass eigenstate representation, the interaction terms are expressed as
\begin{eqnarray}
{\cal L}_{\rm Y} &=&
  -\sum_{f=u,d,\ell} \frac{m_f} {v} \left(\xi_f^h \, h \bar ff + \xi_f^H \, H \bar ff - i\xi_f^A \, A \bar f\gamma_5 f \right) \nonumber\\
 & - &\left[ \frac{\sqrt{2}}{v}  V_{ud} H^+ \bar u \left(m_u \xi_u^A P_L 
+m_d \xi_d^A P_R\right) d +\frac{\sqrt{2}}{v} m_\ell \xi^A_\ell  H^+ \bar \nu P_R \ell + h.c.\right] \, , 
\label{eq:yukawa_int}
\end{eqnarray}
where $f=u$ ($d,\ell$),  $V_{ud}$ is the Cabibbo-Kobayashi-Maskawa matrix element, and we assume summation over the three generations. 
The $\xi_f$ factors in Eq.~(\ref{eq:yukawa_int})  for all the four variants of 2HDM,  as in~\cite{Branco:2011iw}  are given in Table~\ref{tab:z2parity}.
\begin{table*}
	\centering
	\begin{tabular}{|c||cc|ccc|c|c|}
		\hline
		& $Z_2(H_1)$& $ Z_2(Q_L)$ & $Z_2(L_L)$ & $Z_2(H_2)$ & $Z_2(u_R^c)$ & $Z_2(d_R^c)$ & $Z_2(\ell_R^c)$ \\ \hline\hline
		Type-I  & $+$ & $+$ &$+$ & $-$	& $-$ & $-$ & $-$      \\\hline
		Type-II & $+$ & $+$ &$+$ & $-$	& $-$ & $+$ & $+$     
		  \\\hline
		Type-X  & $+$ & $+$ &$+$ & $-$	& $-$ & $-$ & $+$      \\\hline
		Type-Y  & $+$ & $+$ &$+$ & $-$	& $-$ & $+$ & $-$       \\\hline
	\end{tabular}
	\setlength\tabcolsep{2pt}
	\centering
	\begin{tabular}{|c|c|c|c|c|c|c|c|c|c|}
		\hline
		&  $\xi^h_u$ & $\xi^h_d$ & $\xi^h_\ell$
&$ \xi^H_u$ & $\xi^H_d$ & $\xi^H_\ell $ & $\xi^A_u$ & $\xi^A_d$ & $\xi^A_\ell$		
		\\ \hline\hline
		Type-I     &$\cos \alpha/ \sin \beta$  & $\cos \alpha/ \sin \beta$  &$ \cos \alpha/ \sin \beta$
		& $\sin \alpha/ \sin \beta$ & $\sin \alpha/ \sin \beta$ & $\sin \alpha/ \sin \beta$ 
		& $\cot\beta $ & $-\cot\beta$ & $- \cot\beta$ \\\hline
		Type-II   &$\cos \alpha/ \sin \beta$ & $- \sin \alpha/ \cos \beta$ & $- \sin \alpha/ \cos \beta$
		& $\sin \alpha/ \sin \beta$ & $\cos \alpha/ \cos \beta$ & $\cos \alpha/ \cos \beta$
		 & $\cot\beta$ & $\tan\beta$ & $\tan\beta$ \\\hline
		Type-X    & $\cos \alpha/ \sin \beta$ & $\cos \alpha/ \sin \beta$ &$ - \sin \alpha/ \cos \beta$
		& $\sin \alpha/ \sin \beta$ & $\sin \alpha/ \sin \beta$ & $\cos \alpha/ \cos \beta$
		& $\cot\beta$ & $-\cot\beta$ & $\tan\beta$ \\\hline
		Type-Y        & $\cos \alpha/ \sin \beta$ & $- \sin \alpha/ \cos \beta$ & $\cos \alpha/ \sin \beta$
		 & $\sin \alpha/ \sin \beta$ & $\cos \alpha/ \cos \beta$ & $\cos \alpha/ \cos \beta$
		& $\cot\beta$ & $\tan\beta$ & $-\cot\beta$ \\\hline
	\end{tabular}
	\caption{ Assignment of the $Z_2$ parity (top) and $\xi_f$ factors in Eq.~(\ref{eq:yukawa_int}) (bottom) in each type of the Yukawa interactions.}\label{tab:z2parity}
\end{table*}
In our analysis, we have imposed the alignment limit~\cite{Das:2015mwa}
condition on the 2HDM potential which sets the relation between the two CP-even scalar mixing angles as $\sin(\beta -\alpha) = 1$.
In this limit, the tree level couplings of the lightest CP-even state $h$ with mass 125 GeV are exactly the same as the  SM values, 
in agreement with the latest Large Hadron Collider (LHC) Higgs data. This is a simplification, but justified by the consistency of the Higgs data with
the SM predictions, so this is an acceptable approximation for such multi-Higgs doublet scenario~\cite{Das:2019yad}.
Therefore, in the alignment limit, with the lightest Higgs mass $m_h$ set at 125 GeV and the electroweak VEV  $v = 246~\rm GeV$,
there remain only 5 independent parameters in the physical mass basis, which are:  the
three non-SM scalar masses ($m_H\,, m_A\,, m_+$), the soft-symmetry breaking parameter $m_{12}^2$, and  $\tan\beta$.

We now supplement the 2HDM Type-X or Type-II Lagrangian by a single generation of vector-like leptons (VLLs) which includes a set
of leptons with the same quantum numbers as ordinary leptons, and an additional set of mirror leptons with same quantum number
but with opposite chirality. The new VLLs also have the same $Z_2$ charge as the SM leptons. We define the VLLs by the following notation:
\begin{eqnarray}
L_L&=&({\bf 1}, {\bf 2}, -1/2),  \qquad e_{4R}=({\bf 1}, {\bf 1}, -1)\\
L_R&=&({\bf 1}, {\bf 2}, -1/2),  \qquad e_{4L}=({\bf 1}, {\bf 1}, -1)\, ,
\end{eqnarray}
where we have taken the charge  as $Q = T_3 + Y$. 
Hence, within  Type-X or Type-II 2HDM Yukawa assignments, these VLLs couple only to the $H_1$ doublet and the Yukawa
Lagrangian can be written as
	\begin{eqnarray}
	{\cal L}_{\rm Y}^{\rm VLL}&=& 
	- m_{L_4} {\bar L}_L L_R -m_{E_4}{\bar e}_{4L} e_{4R} -
	y_1 {\bar L}_L H_1 e_{4R} - y_2 {\bar L}_R H_1 e_{4L}   + h.c.,
	\label{eq:Yukawa}
	\end{eqnarray}
where $y_1$ and $y_2$ are the VLLs Yukawa couplings and where we assume, for simplicity, that the vector-like leptons and ordinary leptons do not mix. In general, mixing would be allowed with the third family only ($\tau, \nu_\tau)$ to avoid lepton flavor-changing interactions. Such mixing would be in general expected to be small, but yield possibly interesting collider signals  \cite{Ishiwata:2013gma}. The mixing would not affect our considerations of \gmin2 significantly, while introducing additional parameters, and thus we ignore it.

The charged lepton mass matrix for vector-like states, then, takes the form
\begin{equation}
{\cal L} \supset\left( \bar{E_L} \hspace{2mm} \bar{{e_{4L}}}\right) {\cal{M}}
\left( \begin{array}{c}
E_R\\
e_{4R}
\end{array}\right) + h.c. \, ;  \qquad
 {\cal{M}} =  \left( \begin{array}{cc}
 m_{L_4} & \displaystyle \frac{y_1 v_1}{\sqrt{2}} \\
\displaystyle  \frac{y_2^* v_1}{\sqrt{2}}  & m_{E_4}  \\
\end{array}
\right) \, .
\label{eq:massVLL}
\end{equation}
This mass matrix ${\cal{M}}$ can be diagonalized by two bi-unitary transformations $U_L$ and $U_R$, 
${\cal{M}}_D = U_L {\cal{M}} U_R^{\dag}$ yielding  the eigenstates
\begin{equation}
l_L^\prime \equiv
\left( \begin{array}{c}
\ell_{1}\\
\ell_{2}
\end{array}\right)_L = U_L 
\left( \begin{array}{c}
E_L\\
e_{4L}
\end{array}\right)~~ {\rm and}~~
l_R^\prime \equiv
\left( \begin{array}{c}
\ell_{1}\\
\ell_{2}
\end{array}\right)_R= U_{R}
\left( \begin{array}{c}
E_R\\
e_{4R}
\end{array}\right)~,
\label{eq:massdiag}
\end{equation}
with the diagonalizing matrices defined as
\begin{equation}
U_{L(R)}=
\left(\begin{array}{cc}
\cos\theta_{L(R)} & \sin\theta_{L(R)} \\
-\sin\theta_{L(R)} & \cos\theta_{L(R)}
\end{array}\right)\,.
\label{eq:diagmatrices}
\end{equation}
 The spectrum consists of
two mass eigenstates $(\ell_1$, $\ell_2)$ in charged sector with masses $m_{\ell_1,\ell_2}$, 
while the vector-like neutrino mass arises solely from the bare mass term $- m_{L_4} \bar{\nu}_{4L}\nu_{4R} + \text{h.c.}$,  
with the mass eigenstate $m_N = m_{L_4}$. For simplicity we consider  Yukawa coupling with $\Re ( y_1) = \Re (y_2) =y_L$.

The corresponding couplings with the physical Higgs scalars are shown in Table~\ref{tab:yukawa}, 
\begin{table}[tp!]
	\centering
	\begin{tabular}{|c|c|}
		\hline
		Higgs state &  $\displaystyle  y_{X \bar{\ell_i } (\bar N)\ell_i} $ \\ \hline \hline
		$X=h$ & $\displaystyle \frac{1}{\sqrt{2}} \lambda_{h_1 \bar{\ell_i} \ell_i} \cos \beta$ \\ \hline
		$X=H$ & $ \displaystyle \frac{1}{\sqrt{2}} \lambda_{h_1 \bar{\ell_i} \ell_i}  \sin \beta$ \\ \hline
		$X=A$ & $\frac{1}{\sqrt{2}} \lambda_{a_1 \bar{\ell_i} \ell_i} \sin \beta$ \\ \hline
		$X=H^+$ & $\frac{1}{\sqrt{2}} \lambda_{H^+ \bar{N} \ell_{iR(L)}}\sin \beta$\\ \hline
	\end{tabular}
	\caption{Yukawa couplings of VLLs with the physical Higgs scalars in Type-X and Type-II 2HDM in the alignment limit.}
	\label{tab:yukawa}
\end{table}
where the neutral Higgs couplings are
\begin{subequations}
\begin{eqnarray}
-\lambda_{h_1 \bar{\ell_1} \ell_1} &=& y_L (\cos \theta_L \sin \theta_R + \cos \theta_R \sin \theta_L ) + i (\cos \theta_L \sin \theta_R y_{1}^i - \cos \theta_R \sin \theta_L y_2^i)\,,  \\
-\lambda_{h_1 \bar{\ell_2} \ell_2} &=& - y_L (\cos \theta_L \sin \theta_R + \cos \theta_R \sin \theta_L ) + i (\cos \theta_R \sin \theta_L y_{1}^i - \cos \theta_L \sin \theta_R y_2^i)\,,  \\
-\lambda_{a_1 \bar{\ell_1} \ell_1} &=&  y_L (\cos \theta_L \sin \theta_R - \cos \theta_R \sin \theta_L ) + i (\cos \theta_R \sin \theta_L y_{1}^i + \cos \theta_L \sin \theta_R y_2^i)\,, \\
-\lambda_{a_1 \bar{\ell_2} \ell_2} &=&  y_L (\cos \theta_L \sin \theta_R - \cos \theta_R \sin \theta_L ) - i (\cos \theta_R \sin \theta_L y_{1}^i + \cos \theta_L \sin \theta_R y_2^i)\,. 
\label{eq:VLLparam}
\end{eqnarray}
and, the charged Higgs couplings in Table \ref{tab:yukawa} are
\begin{eqnarray}
-\lambda_{H^+ \bar{N} \ell_{1R(2R)}} &=& y_1 ( \sin\theta_R (\cos\theta_R)) \nonumber \\
-\lambda_{H^+ \bar{N} \ell_{2R(2L)}} &=& y_2 ( \sin \theta_L(\cos \theta_L))\,.
\label{eq:VLLparam2}
\end{eqnarray}
\end{subequations}

\section{Constraints on the Parameter Space}
\label{sec:constraints}
Before embarking on the analysis of the effects of VLL on the magnetic moment of muons, we summarize the restrictions on the parameter space, both from experimental constraints on the masses, but also from precision electroweak measurements and theoretical considerations. 

\subsection{Perturbativity, Vacuum Stability and Unitarity}
\label{subsec:theory}

First, perturbativity of all quartic couplings is ensured by imposing the  condition
$|\lambda_i| \leq 4 \pi $.
The requirement of positivity of the potential enforces the following conditions on the quartic
couplings \cite{Sher:1988mj}
\begin{eqnarray}
\lambda_{1,2} > 0 \quad \rm{and} \quad \lambda_3 + \lambda_4 -|\lambda_5| +
\sqrt{\lambda_1 \lambda_2} >0 \quad\rm{and} \quad\lambda_3+ \sqrt{\lambda_1
  \lambda_2} > 0~.
\end{eqnarray}
Finally, we  constrain the model parameters by requiring tree level unitarity for
the scattering of Higgs bosons and longitudinal parts of the electroweak gauge bosons. In 2HDM the necessary
and sufficient conditions for the $S$-matrix to be unitarity in terms of its eigenvalues are derived
in \cite{Ginzburg:2005dt,Horejsi:2005da}. The eigenvalues of $S$-matrix, restricted to $\le 8 \pi$ are given in terms of the couplings in the Higgs potential, Eq. (\ref{eq:higgspotential}), to be
\begin{eqnarray}
&&\lambda_3 \pm \lambda_4 \le 8 \pi \quad , \quad
 \lambda_3 \pm \lambda_5 \le 8 \pi~,\\
&& \lambda_3+ 2 \lambda_4 \pm 3\lambda_5 \le 8 \pi \quad , \quad
\frac{1}{2}\{ \lambda_1 + \lambda_2 \pm \sqrt{(\lambda_1 - \lambda_2)^2 + 4 \lambda_4^2} \} \le 8 \pi~,
\\
&&
 \frac{1}{2}\{ 3\lambda_1 + 3\lambda_2 \pm \sqrt{9(\lambda_1 - \lambda_2)^2 + 4 (2\lambda_3 +
   \lambda_4)^2} \}\le 8 \pi ~,
\\
&&
 \frac{1}{2} \{ \lambda_1 + \lambda_2 \pm \sqrt{(\lambda_1 - \lambda_2)^2 + 4 \lambda_5^2} \} \le 8 \pi~.
\end{eqnarray}
As mentioned, we have considered the {\it alignment limit} condition
of 2HDM potential which renders an exact SM-like coupling for the lightest CP-even Higgs of mass 
125 GeV at tree-level. In such limit, 
the degenerate mass scenario for all the non-SM Higgs bosons naturally
satisfies the unitarity conditions at the electroweak scale~\cite{Bhattacharyya:2013rya,Das:2015mwa} for large $\tan \beta$.

\subsection{VLL mass restrictions from colliders}
\label{subsec:masslim}

Constraints on VLL masses are weak: for sequential charged heavy leptons, $m_{L_4}, m_{E_4}>100.8$ GeV \cite{Tanabashi:2018oca}, while for heavy stable charged leptons the limits are slightly modified $m_{L_4}, m_{E_4}>102.6$ GeV \cite{Tanabashi:2018oca}. The searches for heavy neutral leptons look for  Dirac or Majorana fermions with sterile neutrino quantum numbers,  heavy enough not to disrupt the simplest Big Bang Nucleosynthesis bounds and/or stability on cosmological timescales.  Mass limits are ${\cal O}$(MeV) or higher \cite{Tanabashi:2018oca}. Rather than establishing firm mass limits, searches for these particles generically set bounds on the mixing between them and the three SM neutrinos \cite{Tanabashi:2018oca}, which are not applicable here, as we neglect them. In what follows, we shall assume, allowing for uncertainties, that for all VLLs, $m_{l_1}, m_{l_2}, m_N \ge 100$ GeV.

\subsection{Effects of VLLs in Higgs diphoton decays}
\label{subsec:hgg}

Our assumption of the alignment limit renders exact SM-like tree-level couplings for $h$ with fermions and vector bosons.
The collider mass restriction on the VLLs, as discussed in Sec.~\ref{subsec:masslim}, will only modify three body
decay modes for $h$, changing the total decay width negligibly. However, the charged VLLs ($l_1,l_2$) can contribute to the loop-induced
decay mode of the Higgs into $\gamma \gamma$ along with
the charged scalar $H^\pm$ of the 2HDM scalar potential. The current experimental limit on the Higgs to 
diphoton signal strength is quite close to its SM value and stands at $\mu_{\gamma\gamma} \equiv \frac{\mu_{\gamma\gamma}^{\rm exp}}{\mu_{\gamma\gamma}^{\rm SM}}= 1.18^{+0.17}_{-0.14}$ ~\cite{Sirunyan:2018ouh}.
Therefore, it is mandatory to require our model to be consistent with the current Higgs to diphoton decay limit. 
The  Higgs to diphoton decay width is expressed in terms of the couplings
to the particles in the loop as 
\begin{equation}
\Gamma (h \to \gamma \gamma) = \displaystyle \frac{\alpha^2\, g^2
	}{1024 \pi^3}\frac{m_{h}^3}{M_W^2} \left \vert F_{1} (x_W) + \frac{4}{3}
F_{1/2}(x_t) + \sum_{i=1}^{2}  y_{h \bar{\ell_i} \ell_i } F_{1/2}(x_{\ell_i} ) + \kappa_{hH^{+}H^{-}}
F_+(x_{H^+}) \right \vert^2  \, ,
\label{eq:hgg}
\end{equation}
where  $x_j \equiv (2 m_j/ m_h)^2$, $j=W,t,f,H^+$, $m_{h}$ is the SM Higgs
mass,  and $y_{h \bar{\ell_i} \ell_i }$} ($\kappa_{hH^{+}H^{-}}$) are
the couplings of SM Higgs boson to vector-like leptons (charged Higgs) with
mass  $m_{\ell_i}$ ($m_{+}$), respectively. The loop functions $F_{1},
F_{1/2}$ and $F_+$  which  appear in the  calculation of decay width
$\Gamma(h \to \gamma\gamma)$ are~\cite{Gunion:1989we}:
\begin{eqnarray}
&& F_1(x) = 2 + 3 x + 3x(2-x)f(x),\nonumber\\
&& F_{1/2}(x)= -2 x[1+(1-x)f(x)],\nonumber\\
&& F_+(x) = -x[1-xf(x)]
\end{eqnarray}
with
\begin{equation}
f(x) = \left\{ \begin{array}{lr}
[\sin^{-1}(1/\sqrt{x})]^2, & x \geq 1 \\
-\frac{1}{4} [\ln\left(\frac{1 + \sqrt{1-x}}{1 - \sqrt{1-x}}\right) - i \pi]^2, & \, x < 1
\end{array}  \right.
\end{equation}
and, the charged Higgs couplings to the SM Higgs is given by~\cite{Djouadi:1996yq}
\begin{eqnarray}
\kappa_{hH^{+}H^{-}} &=&- \frac{1}{2\, m_+^2}  \left(2 (m_h^2 - m_0^2) \frac{\cos(\beta + \alpha)}{
\sin 2 \beta} + \left(2 m_+^2 - 
m_h^2) \sin(\beta - \alpha)\right)\right),   
\end{eqnarray}
where   $m_0^2=\frac{m_{12}^2}{\sin \beta \cos \beta}$.
In the SM  the loop contributions to $h \to \gamma \gamma$  come from the top quark and $W$ gauge boson circulating in the loop,  with a loop
factor  of
$F_1(x_W) \to + 8.3$ and  $F_{1/2}(x_t) \to -
1.34$ for $m_h=125$\,GeV. Since VLLs do not contribute to Higgs production, we define the
ratio of decay width describing the enhancement/suppression in $h\rightarrow \gamma\gamma$ channel
\begin{equation}
\mu_{\gamma\gamma} = \frac{\sigma(pp \rightarrow h)}{\sigma_{SM}(pp \rightarrow h)}\frac{\Gamma (h \to \gamma \gamma)}{\Gamma_{SM} (h \to \gamma \gamma)}
=\frac{\Gamma (h \to \gamma \gamma)}{\Gamma_{SM} (h \to \gamma \gamma)}~.
\end{equation}
In Fig. \ref{fig:hgg}, we illustrate the restrictions on mass parameters $m_0\,, m_L\,,m_+$ for ${\cal O}(1)$ Yukawa couplings and fixed $\tan\beta =10$, and for a particular choice of VLLs mixing angles $(\sin\theta_L=0.5\,,\sin\theta_R = -0.4)$ from the requirement that $\mu_ {\gamma \gamma}$ agrees with the experimental value at $2\sigma$. 
In the left panel of  Fig. \ref{fig:hgg}, we show the allowed region for two different choices of charged Higgs masses, $m_+ = 200$ and 300 GeV,  drawn in 
orange and green shades respectively. As one can see, for a particular charged Higgs mass, $m_0$ can  be at most (a little more than) twice as large as $m_+$ for any values of $m_L$. While on the right panel  of Fig. \ref{fig:hgg}, we fixed both the charged VLL mass at $m_L=150$ GeV 
and show the allowed region in $m_+ - m_0$ plane. It is clear from these graphs that a proper tuning between the soft-symmetry breaking term $m_0$, the charged Higgs mass $m_+$ and the VLL mass parameter $m_L$ can easily satisfy the experimental data for $h \to \gamma \gamma$. It is to be noted that although we chose a particular value for the VLL Yukawa coupling and $\tan\beta$, the correlation between the mass parameters does not depend crucially on the choice. The reason behind the choice of the VLLs mixing angles relies on our later analysis of oblique parameter constraints and muon anomalous magnetic moment values, as we will discuss in the following sections.
\begin{figure}[htbp!]
	\centering
	\includegraphics[width=80mm, height=80mm]{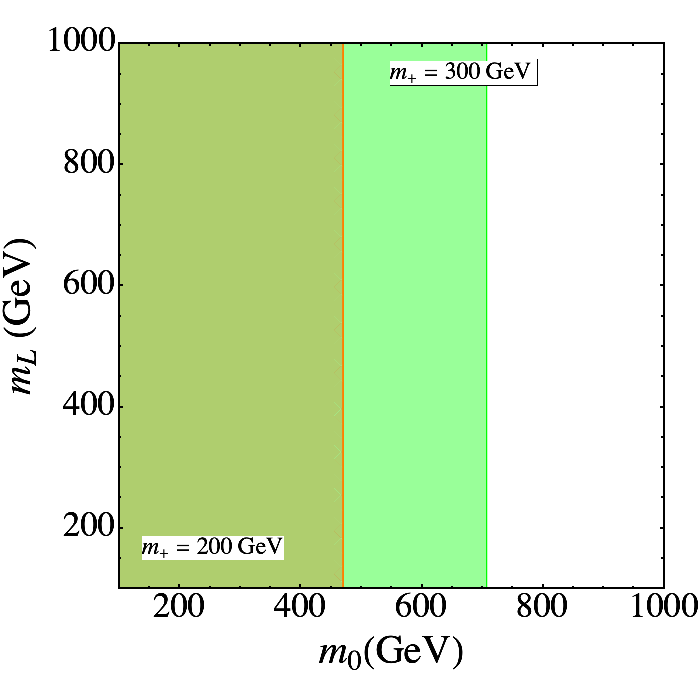} 
	\includegraphics[width=80mm, height=80mm]{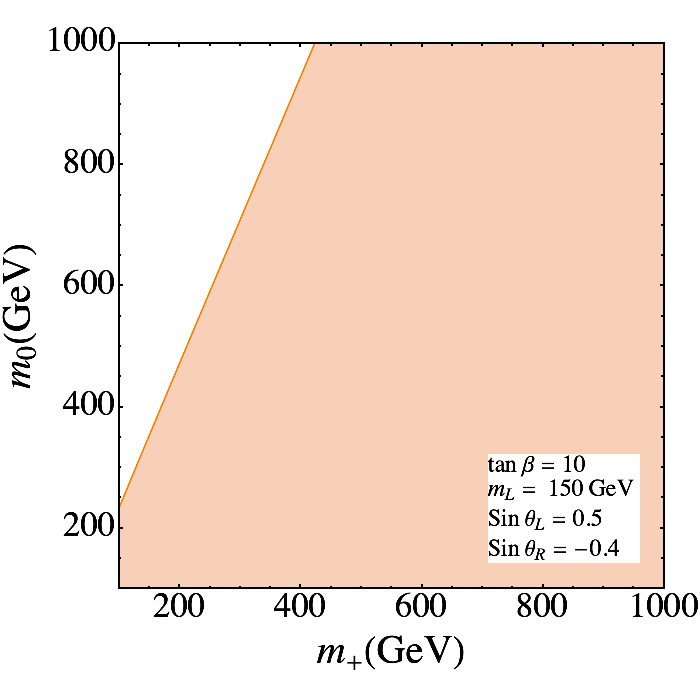} 
	\caption{ Restriction on ($m_0$ - $m_L$) (left) and ($m_0$ - $m_{+}$) (right) plane from imposing the $2\sigma$ limit of $h \to \gamma \gamma$
		signal strength. Here $m_0$ is the soft breaking parameter defined in the text, and $m_L$ is the degenerate mass for the two vector-like charged leptons. In the left panel, the two shaded regions (orange and green) correspond to two different choices of charged Higgs mass,
	$m_+ = 200$ and 300 GeV respectively. The choice of $\tan\beta$, Yukawa coupling and VLL mixing angle, given in the right-hand side panel, is same for both the figures.}
		\label{fig:hgg}
\end{figure}

\subsection{Gauge Boson Couplings and Oblique Parameters}
\label{subsec:oblique}
In addition to the Higgs data and the theoretical constraints defined in the previous subsections,
 crucial restrictions come from the electroweak oblique parameters, as the 
additional scalars and leptons  contribute to gauge boson masses via loop corrections.
The scalar contributions to the oblique $T$ and $S$ parameters are well-known and can be found in~\cite{Grimus:2007if,Grimus:2008nb}.

For the VLLs contribution, we first compute the VLL gauge couplings with the vector bosons, 
with the VLL mass matrices defined by Eq.~(\ref{eq:massdiag}), and with the components
of the diagonalizing matrices  given by Eq. (\ref{eq:diagmatrices}). The $W$ couplings with VLLs can be written as,
\begin{eqnarray}
{\cal L}_W &=& \frac{g}{\sqrt{2}} \left({\bar f}_L^i \gamma^\mu A_{ij}^L f_L^j + {\bar f}_R^i \gamma^\mu A_{ij}^R f_R^j \right)W_\mu^+ +  h.c. \nonumber \\
&=& \frac{g}{\sqrt{2}} {\bar N} \gamma^\mu \left(A_{i}^L l_L^i + A_{i}^R l_R^i  \right)  W_\mu^+ +  h.c. \, ,
\end{eqnarray}
where the couplings are given by,
\begin{eqnarray}
A_{1}^{L(R)} = (\cos \theta^{L(R)}) \,, \qquad
A_{2}^{L(R)} = (-\sin \theta^{L(R)}) \,. 
\label{Aij}
\end{eqnarray}
The neutral $Z$-boson couplings with VLLs are given by
\begin{eqnarray}
{\cal L}_Z &=& \frac{g}{2 \cos\theta_W} \left({\bar f}_L^i \gamma^\mu c_z f_L^j + {\bar f}_R^i \gamma^\mu c_z  f_R^j \right)Z_\mu  \, \nonumber \\
&=&   \frac{g}{2 \cos\theta_W} \left({\bar \ell}_i^L \gamma^\mu X_{ii}^L\ell_i^L  + {\bar \ell}_i^R \gamma^\mu X_{ii}^R  \ell_i^R- \frac{1}{2}{\bar N}\gamma^\mu N  \right)Z_\mu \, ,
\label{eq:Zcouplings}
\end{eqnarray}
where $c_z =(T_3 - Q \sin^2\theta_W)$ and the values of $X_{ii}$  are  
\begin{eqnarray}
X_{11}^L=X_{11}^R=\frac{1}{2}(2\sin^2\theta_W -1 )\, , \qquad  X_{22}^L=X_{22}^R=\sin^2\theta_W\, .
\end{eqnarray}
%
We now proceed to analyze separately the new contributions of VLLs to the  $T$ and $S$ parameters. 
\subsubsection{$T$-parameter}
\label{subsubsec:T}
The general expression for the $T$-parameter contribution from additional fermions is \cite{Chen:2017hak},
\begin{eqnarray}
\Delta T_F &=& \frac{1}{8 \pi s^2_w c^2_w} \sum_{i,j} \big[(|A_{ij}^L|^2 + |A_{ij}^R|^2)\theta_+ (f_i, f_j) + 2 \Re(A_{ij}^L A_{ij}^{R*}) \theta_- (f_i, f_j)  \nonumber \\
&& -\frac{1}{2}\left( (|X_{ij}^L|^2 + |X_{ij}^R|^2)\theta_+ (f_i, f_j) + 2 \Re(X_{ij}^L X_{ij}^{R*}) \theta_- (f_i, f_j) \right) \big]
\end{eqnarray}
where, $f_i = \frac{m_{f_i}^2}{M_Z^2}$ and the functions are defined as,
\begin{subequations}
\begin{eqnarray}
\theta_+\left(x,y
\right)&=&\left\{
\begin{array}{cc}
\displaystyle \frac{x+y}{2} - \frac{xy}{x-y}\ln\left(\frac{x}{y} \right) &  x\neq y \,, \nonumber \\
0 & x=y \,.
\end{array}
\right. \\
\theta_-\left(x,y
\right)&=&\left\{
\begin{array}{cc}
\displaystyle \sqrt{xy}\left[ \frac{x+y}{x-y}\ln\left(\frac{x}{y} \right) -2\right] &  x\neq y \,, \nonumber \\
0 & x=y \,.
\end{array}
\right. 
\end{eqnarray}
\end{subequations}
\subsubsection{$S$-parameter}
\begin{figure}[htbp!]
	\centering
	\includegraphics[width=80mm, height=80mm]{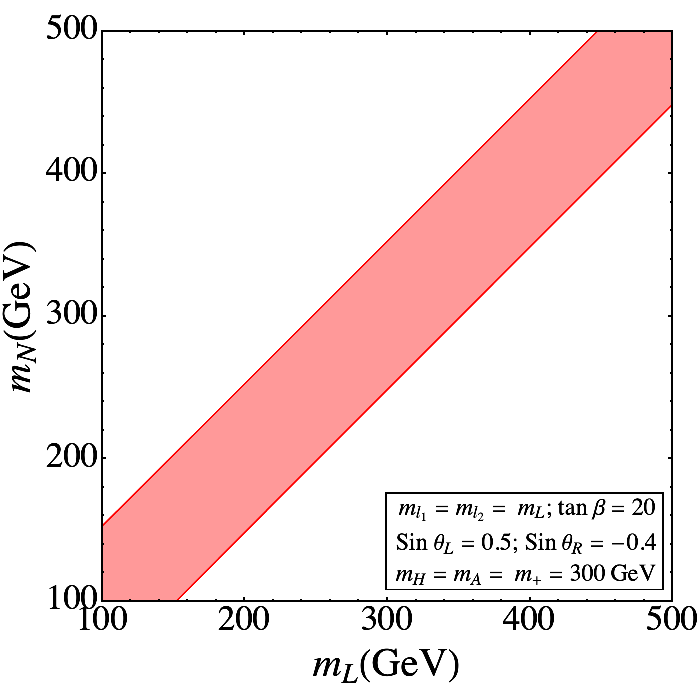} 
	\includegraphics[width=80mm, height=80mm]{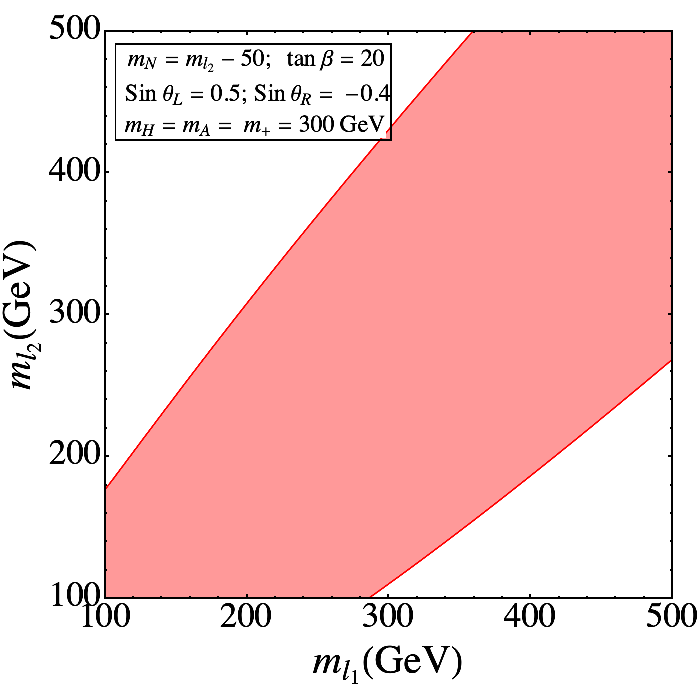} \\
	 	\includegraphics[width=80mm, height=80mm]{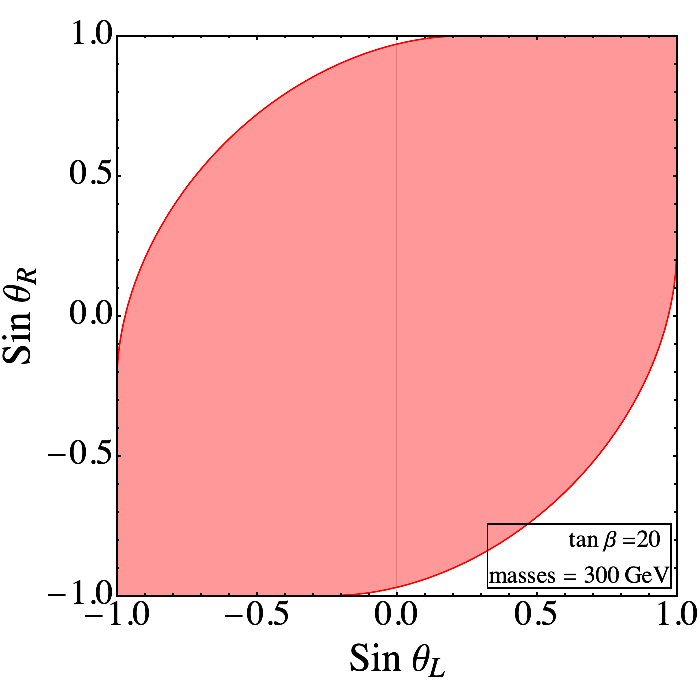} 
	 \includegraphics[width=80mm, height=80mm]{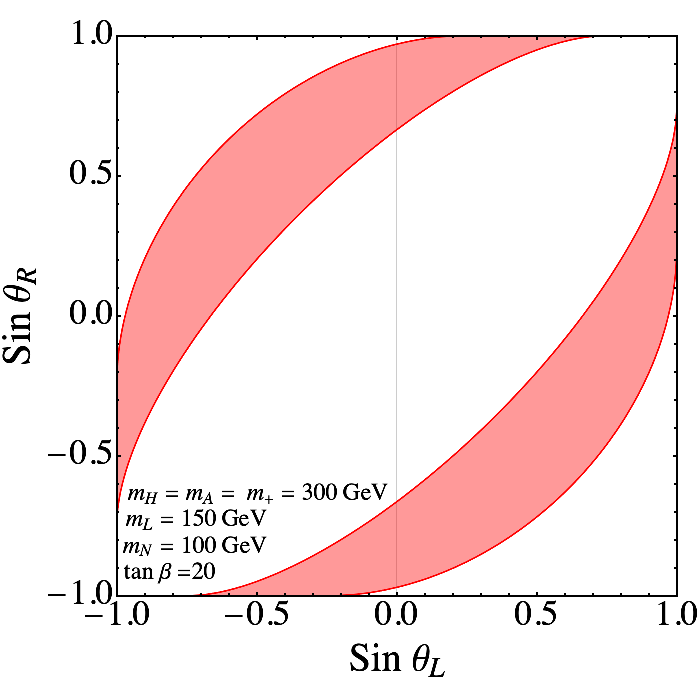}
	\caption{ Constraints on the $S$ and $T$ parameters as functions of masses and mixing parameters of VLL. Upper Panel: Constraints in the neutral and charged VLL masses in the $(m_N-m_L)$  and  $(m_{l_1}-m_{l_2})$ plane. Lower Panel: Constraints on the mixing angles in the left- and right-handed VLL sector in the $\sin(\theta_L)-\sin(\theta_R)$ plane for two different benchmark scenarios. (Left) Degenerate VLL masses,  $m_{l_1}=m_{l_2} = m_N = $ 300 GeV; (Right) $m_{l_1}=m_{l_2} \equiv m_L= $ 150 GeV, $m_N=100$ GeV.  We take throughout $\tan \beta =20$ and degenerate non-standard Higgs masses are fixed at 300 GeV. }
	\label{fig:ST}
\end{figure}

The general expression for the $S$-parameter contribution from additional fermions is \cite{Chen:2017hak},
\begin{eqnarray}
\Delta S_F &=& \frac{1}{2 \pi} \sum_{i,j} \big[(|A_{ij}^L|^2 + |A_{ij}^R|^2)\psi_+ (f_i, f_j) + 2 \Re(A_{ij}^L A_{ij}^{R*}) \psi_- (f_i, f_j)  \nonumber \\
&& -\frac{1}{2}\left( (|X_{ij}^L|^2 + |X_{ij}^R|^2)\chi_+ (f_i, f_j) + 2 \Re(X_{ij}^L X_{ij}^{R*}) \chi_- (f_i, f_j) \right) \big]
\end{eqnarray}

where, as before, $f_i = \frac{m_{f_i}^2}{M_Z^2}$ and the functions are defined as,
\begin{subequations}
	\begin{eqnarray}
	\psi_+\left(x,y\right) &=& \frac{1}{3} - \frac{1}{9}\ln\left(\frac{x}{y}\right) \\
	\psi_-\left(x,y\right) &=&  - \frac{x+y}{6\sqrt{xy}} \\
	\chi_+\left(x,y
	\right)&=&\left\{
	\begin{array}{cc}
	\displaystyle \frac{5(x^2+y^2)-22xy}{9(x-y)^2} + \frac{3xy(x+y)-x^3 - y^3}{3(x-y)^3}\ln\left(\frac{x}{y} \right) &  x\neq y \,, \nonumber \\
	0 & x=y \,.
	\end{array}
	\right. \\
	\chi_-\left(x,y
	\right)&=&\left\{
	\begin{array}{cc}
	\displaystyle -\sqrt{xy}\left[ \frac{x+y}{6xy} - \frac{x+y}{(x-y)^2} +\frac{2xy}{(x-y)^3} \ln\left(\frac{x}{y} \right)\right] &  x\neq y \,, \nonumber \\
	0 & x=y \,.
	\end{array}
	\right. 
	\end{eqnarray}
\end{subequations}
Inserting the couplings from Eq.~(\ref{Aij}), we obtain
\begin{eqnarray}
\Delta T_F &=& \frac{1}{4 \pi s^2_w c^2_w} \bigg(2\theta_+(f_1,f_N)+ ({s_\theta^L}^2 + {s_\theta^R}^2)[-\theta_+(f_1,f_N) + \theta_+(f_2,f_N)] \nonumber \\
&& 2(c_\theta^L c_\theta^R )[\theta_-(f_1,f_N)] + 2(s_\theta^R s_\theta^L)[\theta_-(f_2,f_N)]  \bigg)\,, \\
\Delta S_F &=& \frac{1}{\pi } \bigg(2\psi_+(f_1,f_N) + ({s_\theta^L}^2 + {s_\theta^R}^2)[-\psi_+(f_1,f_N) + \psi_+(f_2,f_N)] \nonumber \\
&& 2(c_\theta^L c_\theta^R)\psi_-(f_1,f_N) + 2(s_\theta^R s_\theta^L)\psi_-(f_2,f_N)  \bigg)\,.
\end{eqnarray}
The current global electroweak fit yields~\cite{Tanabashi:2018oca}
\begin{eqnarray}
\Delta T &=& 0.07\pm0.12 \,, \qquad  \Delta S = 0.02\pm0.07 \, ,
\end{eqnarray}
which should be satisfied by the total contribution from both the scalars and the new VLLs contributions. In Fig. \ref{fig:ST},
we show the restrictions on the mass and mixing angles of VLL, resulting from imposing restrictions on the $S$ and $T$ parameters at $2\sigma$ uncertainty while keeping all the non-SM scalars masses at the same value, 300 GeV.  A
degenerate mass spectrum for the scalars will not have any impact on the oblique parameters and thus 
the restriction on oblique parameters will impose the relations among  VLL parameters which can be further explored.
As can be seen from the upper panel of Fig. \ref{fig:ST},  for  degenerate charged VLL masses, a mass splitting of 
$(m_L - m_N) \lesssim$ 100 GeV with the vector-like neutrino (VLN) is required to satisfy the oblique corrections. On the other hand, fixing the
mass splitting between the VLN with one of the charged VLLs at 50 GeV  allows  mass splitting larger than 100 GeV
between the two charged VLLs. These results are independent of the choice of VLL mixing angle. 
In the lower panel of Fig. \ref{fig:ST}, we show the correlation between the mixing angles in the left and right-handed sectors, $\theta_L$ and $\theta_R$, restricted by the oblique parameters values at 2$\sigma$. As expected, for a complete degenerate mass scenario
requiring the non-SM scalars and the VLLs to have same mass, a larger parameter region is allowed by the $S$ and $T$ parameters (bottom left), where the degenerate masses are  $m_L=m_H=m_A=300$ GeV.  Note that small changes in the degenerate VLL mass does not alter the left panel significantly. 
However, fixing the VLL lepton masses at $m_{l_1}=m_{l_2} \equiv m_L= $ 150 GeV, with the VLN mass at $m_N=100$ GeV,
a significant reduction in the allowed parameter space can be seen (bottom right). In view of these analyses, we fix the VLL mixing angles at $\sin \theta_L = 0.5 $ and $\sin\theta_R = -0.4$ for the following analysis on \gmin2, as a viable choice, which is also compatible with \gmin2 as we will discuss further. Finally, we note that the mass and mixing angle correlation is independent of our choice of $\tan\beta$ and of the degenerate scalar mass.

\section{Muon anomalous magnetic moment with VLL}
\label{sec:g-2}

\begin{figure}[htbp!]
	\centering
	\includegraphics[width=40mm]{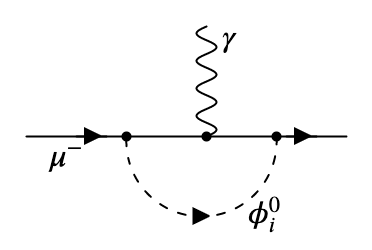}
	\includegraphics[width=40mm]{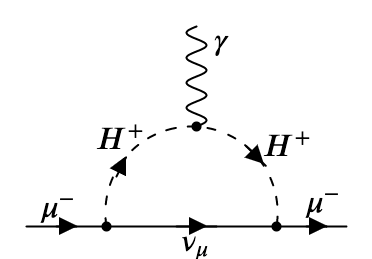}
	\caption{One loop diagrams contribution to magnetic moments in 2HDM. Here $\phi_i^0=h, H,A$.}
	\label{fig:oneloop}
		\end{figure}

\begin{figure}[htbp!]
	\centering
	\begin{tabular}{cccc}
	\includegraphics[width=30mm]{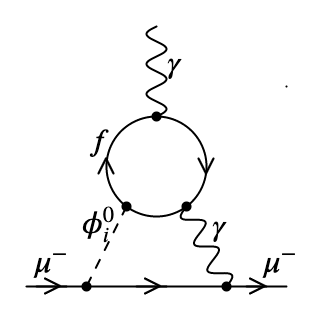} &
	\includegraphics[width=30mm]{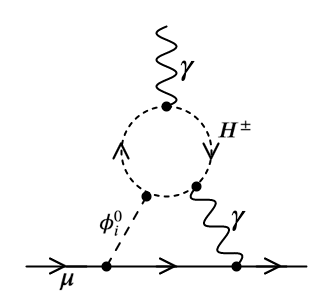} &
	\includegraphics[width=30mm]{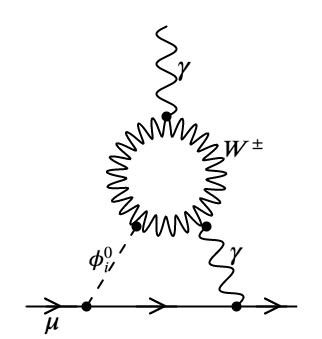} &
	\includegraphics[width=30mm]{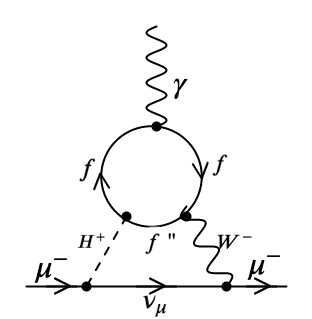} \\
	(a) & (b) & (c) & (d)
		\end{tabular}
   \caption{Dominant two-loop Barr Zee diagrams in 2HDM with VLLs. Here $f=t, b,\ell$ and $f^{\prime\prime}=b, t,N$ where $\ell= \ell_1, \ell_2$, and  $\phi_i^0=h, H,A$.}
   \label{fig:BarrZee}
\end{figure}

We now discuss how the Two Higgs Doublet Model with one VLL generation resolves the long-standing anomaly of the muon magnetic
moment.   In the SM,  the muon magnetic moment originates from the one-loop contributions with diagrams with the Higgs, $Z$ and $W$ bosons and it has been studied extensively in the past~\cite{Leveille:1977rc,Lynch:2001zs,Kannike:2011ng,Jegerlehner:2009ry}. The one-loop contribution to 
the \gmin2, in a 2HDM regime, is supplemented by additional neutral and charged Higgs loops, as depicted in Fig.~\ref{fig:oneloop},  and it has been analyzed in the context of the Type-X 2HDM scenario~\cite{Abe:2015oca}. In the 2HDM,  the \gmin2  contributions are further
enhanced by two-loop Barr Zee diagrams contributing
 with similar strength as the one-loop diagrams~\cite{Chang:2000ii,Cheung:2001hz,Arhrib:2001xx,Krawczyk:2002df,Cheung:2003pw,Heinemeyer:2003dq,Broggio:2014mna,Wang:2014sda}, as the two-loop contributions have a loop suppression factor of  $(\alpha/\pi)$ but benefit from an enhancement of $(M^2/m_\mu^2)$, with $M$ the mass of the heavy particle in the loop. A list of all relevant two-loop Barr Zee diagrams have been given  in~\cite{Ilisie:2015tra} together with the corresponding analytical expressions.
 In our case, the most relevant additional Barr Zee diagrams are shown in Fig. \ref{fig:BarrZee}, where the fermion loop in Fig. \ref{fig:BarrZee}(a) includes the charged VLLs. Moreover,  Fig. \ref{fig:BarrZee}(d) will be non-negligible for the VLLs, while  for ordinary leptons it is suppressed by the small neutrino mass. Similarly, in Fig. \ref{fig:BarrZee}(d), one should also add a diagram where the VLL loop is replaced by the top and bottom quarks, but in Type-X 2HDM, the quark couplings are suppressed by $\cot\beta$. However, for Type-II 2HDM, they could contribute non-negligibly since the bottom quark Yukawa is also proportional to 
 $\tan\beta$. We have, however, included all the contributions in our analysis. The analytical expressions for the one-loop diagrams appeared in \cite{Chun:2016hzs}. As we neglect mixing between ordinary and vector-like leptons, we do not have additional contributions arising from one-loop diagrams.
 
The analytical expression for the additional Barr Zee type diagrams including the VLL loops  is:
\begin{subequations}
\begin{eqnarray}
\Delta a_\mu^{\rm 2HDM} ({\rm Fig.}~\ref{fig:BarrZee}}~{\rm a,b,c) &=&  \frac{\alpha_{\rm EM}  m_\mu^2}{32 \pi^3 v^2} \left[\sum_{i=h,H,A} \sum_f N_C^f Q_f^2
y_f^i y_\mu^i G_i\left(\frac{m_f^2}{m_i^2}\right) + 
\sum_{i=h,H} y_\mu^i g_V^i G_W\left(\frac{M_W^2}{m_i^2}\right) \right] \, \nonumber \\
 &&+\frac{m_\mu^2}{16 \pi^2 m_{+}^2} \frac{\alpha_{\rm EM}}{\pi} \sum_{i=h,H} y_\mu^i  y_{H^+ \bar{N} \ell_i}  G_+ \left(\frac{ m_{+}^2}{m_i^2}\right)
\end{eqnarray}
where, $f$ includes both the SM fermions $(t,b,\tau)$ and VLLs $\ell_1,\ell_2$.
\begin{eqnarray}
G_{h,H}(r) &=& r \int_{0}^{1} dx \frac{2 x(1-x)-1}{x(1-x)-r} \ln\left[\frac{x(1-x)}{r}\right] \,, \nonumber \\
G_A(r) &=&  r \int_{0}^{1} dx \frac{1}{x(1-x)-r} \ln\left[\frac{x(1-x)}{r}\right] \,, \nonumber \\
G_W(r) &=&  \int_{0}^{1} dx \frac{x(3x(4x-1)+10)r - x(1-x)}{x(1-x)-r} \ln\left[\frac{x(1-x)}{r}\right] \,, \nonumber \\
G_+(r) &=&  \int_{0}^{1} dx \frac{x(x-1)}{x(1-x)-r} \ln\left[\frac{x(1-x)}{r}\right] \,.
\end{eqnarray}
\end{subequations}
For diagram (d) in Fig.~\ref{fig:BarrZee} with VLL, the contribution is
\begin{subequations}
\begin{eqnarray}
\Delta a_\mu^{\rm 2HDM}(VLL) (\rm Fig.~\ref{fig:BarrZee}~d) &=& \ \frac{\alpha_{\rm EM}  m_\mu^2}{32 \pi^3 v^2 \sin^2 \theta_W} \frac{1}{  ( m_{+}^2 - M_W^2)} \times\, \nonumber \\
&&  \sum_{f = l_1,l_2}  \int_{0}^{1} dx \, Q_L (1- x) \left[\tan \beta \,  y_{H^+ \bar{N} \ell_i}  m^2_f x(1-x) + \tan \beta \,  y_{H^+ \bar{N} \ell_i}  m^2_N x(1+x) \right] \, \nonumber \\ 
&& \times \left[F\left(\frac{m_N^2}{m_{+}^2}, \frac{m_f^2}{m_{ +} ^2} \right) - F\left(\frac{m_N^2}{M_{W}^2}, \frac{m_L^2}{M_W^2}\right) \right] \, 
\end{eqnarray}
where $Q_L= -1$ and
\begin{eqnarray}
F[r_1,r_2] &=& \frac{\ln\left[\frac{r_1 x + r_2 (1-x)}{x(1-x)}\right]}{x(1-x)- r_1 x - r_2(1-x)}\,.
\end{eqnarray}
\end{subequations}
Here, $m_L$ and $m_N$ correspond to the mass of the two (degenerate) charged eigenstates of VLLs, and the vector-like neutrino, respectively, and 
$ y_{H^+ \bar{N} \ell_i}  $ is the charged Higgs coupling with the VLLs, as defined in Table~\ref{tab:yukawa}.
While for diagram Fig.~\ref{fig:BarrZee}(d) with top-bottom loop, the contribution can be found in~\cite{Ilisie:2015tra}.  Note that $(g-2)_\mu$ is real, and while the couplings  $y_{H^+ \bar{N} \ell_i}, y_{h(H)  \bar{\ell}_i \ell_i},   y_{A  \bar{\ell}_i \ell_i}$ can be complex, their imaginary parts are assumed to be small, and thus they will not affect our numerical calculations.

With the above setup, we proceed to perform our analysis of \gmin2. 

\section{Results}
\label{sec:results}

In this section, we analyze our results and show the effects of adding  VLLs to 2HDM. We first analyze  Type-X 2HDM, which was the most promising scenario without VLLs, and show that we are able to considerably enlarge the pseudoscalar mass-$\tan \beta$ parameter space that can
explain the \gmin2 experimental result. To quantify our considerations, we choose some benchmark points for the model parameters,  which satisfy all
coupling constraints, Higgs to diphoton data,  and the oblique parameter constraints, as  discussed in the
previous sections. 
\begin{figure}[htbp!]
	\centering
	\hspace{-1.5cm}
	\includegraphics[width=60mm, height=70mm]{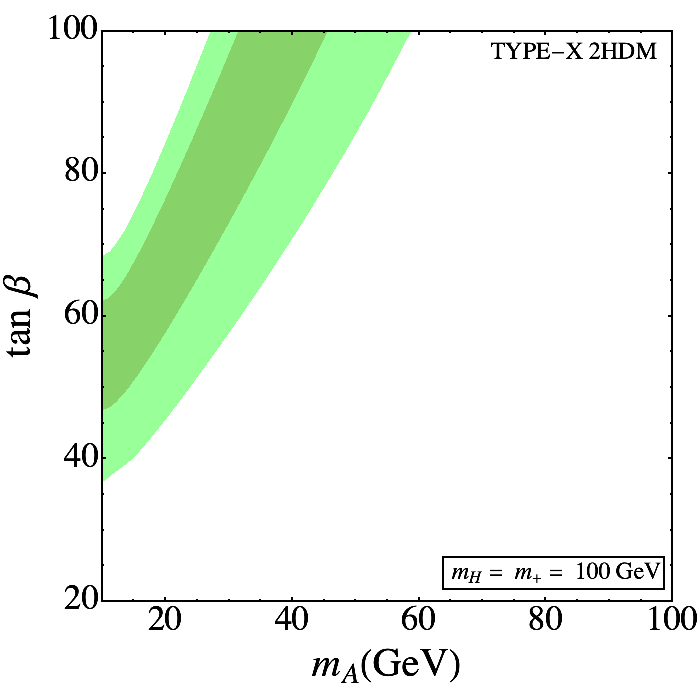} 
	\includegraphics[width=60mm, height=70mm]{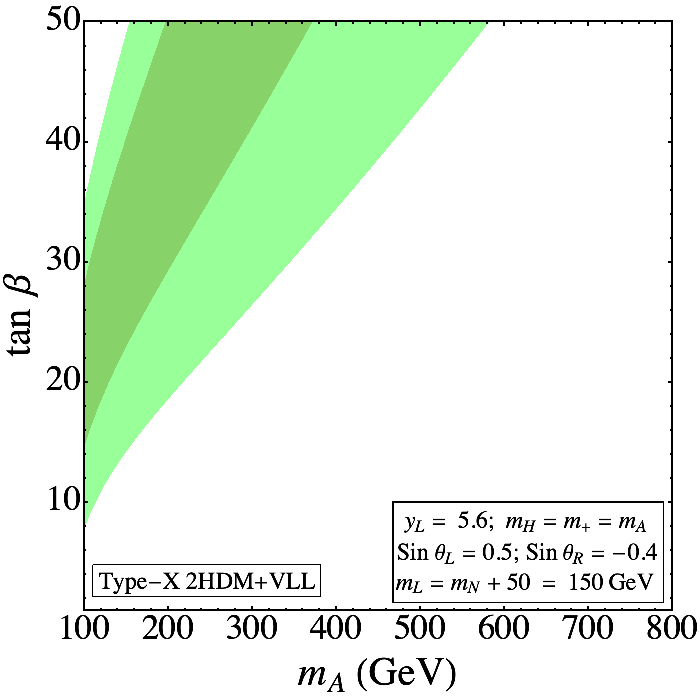} 
	\includegraphics[width=60mm, height=70mm]{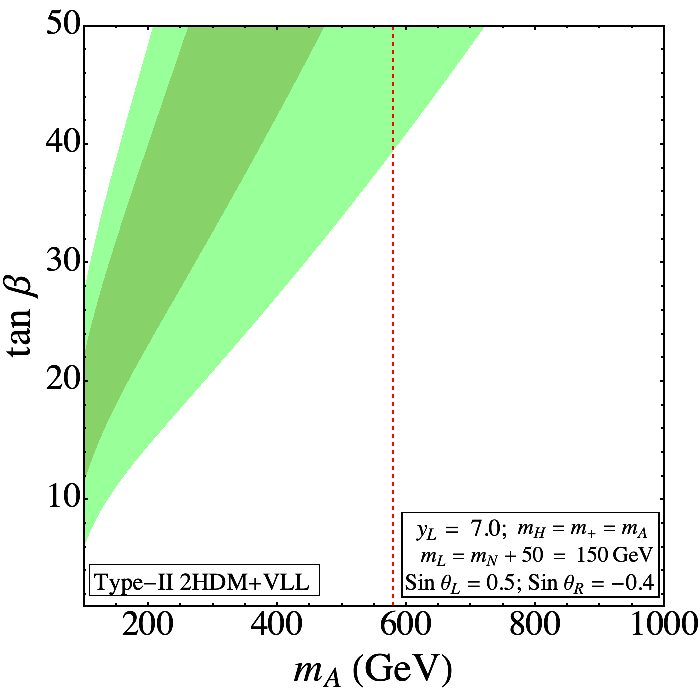}
	\caption{ Comparison of the allowed parameter space in $m_A-\tan \beta$ plane for solutions reproducing the correct value for the muon anomalous magnetic moment in Type-X 2HDM without VLLs (left),  with VLLs (middle) and Type-II 2HDM with VLLs (right). We show the constraints imposed by agreement with the muon $g-2$ at 1$\sigma$ (dark green) and 2$\sigma$ (light green). }
	\label{fig:muMM}
\end{figure}

In Fig.~\ref{fig:muMM}, we show the comparison between the allowed parameter space in $\tan \beta-m_A$ plane
for the Type-X 2HDM scenario without (left panel) and with the extra VLL generation (middle panel). As is evident from the figure,
the introduction of the VLL generation  enlarges significantly the parameter space allowing $m_A$ as large as 600 GeV at $2\sigma$ for 
$\tan\beta \sim 50$,  while without VLLs  Type-X 2HDM can only allow $m_A < 60 $ GeV at a much larger $\tan\beta \sim 100$.  The
dark and light shades of color in the figure refer to the 1$\sigma$ and 2$\sigma$ uncertainty in the \gmin2, as in Eq.~(\ref{eq:gmin2}). 

Moreover the analysis for a Type-II 2HDM with VLLs yields a similar result for \gmin2, validating same parameter range as Type-X.   As mentioned earlier, the bottom Yukawa coupling has the same $\tan\beta$ dependence as the charged leptons in the Type-II 2HDM extension. Therefore,
the contribution from the bottom quark loop in Fig~\ref{fig:BarrZee}(a) and top-bottom loop in Fig~\ref{fig:BarrZee}(d) should not be neglected. But, the bottom quark mass is much less than the VLL mass and the top mass Yukawa coupling is suppressed, so the VLL loop yields the dominant contribution resulting in 
same parameter region as allowed by \gmin2 data in Type-X scenarios. 
However, for Type-II 2HDM, charged Higgs masses $m_+ < 580$ GeV are disallowed at 95\% C.L. from the
BR($B \to X_s \gamma$)~\cite{Belle:2016ufb,Misiak:2017bgg} leaving less parameter space for the  \gmin2 explanation. In the right-hand panel of 
 Fig.~\ref{fig:muMM}, we show an example plot for the allowed parameter space in the $m_A-\tan\beta$ plane for Type-II 2HDM with an extra VLLs model scenario, with 
the color convention  the same as in the other panels. The dashed red line denotes the limit from BR($B \to X_s \gamma$) on the charged Higgs mass, which is taken degenerate with the pseudoscalar mass. The left side of the dashed red line is excluded at 95\% C.L. and the remaining region can satisfy the \gmin2 only at $2\sigma$.
Nevertheless, the VLL extension allows the Type-II 2HDM,  as a valid 2HDM extension which correctly satisfies the \gmin2 anomaly, while without VLLs the whole scenario is ruled out.
In the VLL augmented scenarios, we consider degenerate masses for the non-SM scalars while in the plot on the left panel of Fig.~\ref{fig:muMM}, the heavier Higgs mass is fixed at $m_H=100$ GeV.
The degenerate mass for the non-SM scalars is a more conservative choice, imposed to obey the $S, T, U$ parameter restrictions. 
Our results in the left panel of Fig.~\ref{fig:muMM}, agree with the results of \cite{Chun:2019oix}. In the middle and right panels of Fig.~\ref{fig:muMM}, we take the charged VLL mass fixed at 150 GeV while the VLN mass is 100 GeV. To understand the choice of
VLL mixing angle, we explore the effect on the other model parameters.

\begin{figure}[htbp!]
	\centering
	\begin{tabular}{cc}
		\includegraphics[width=80mm, height=80mm]{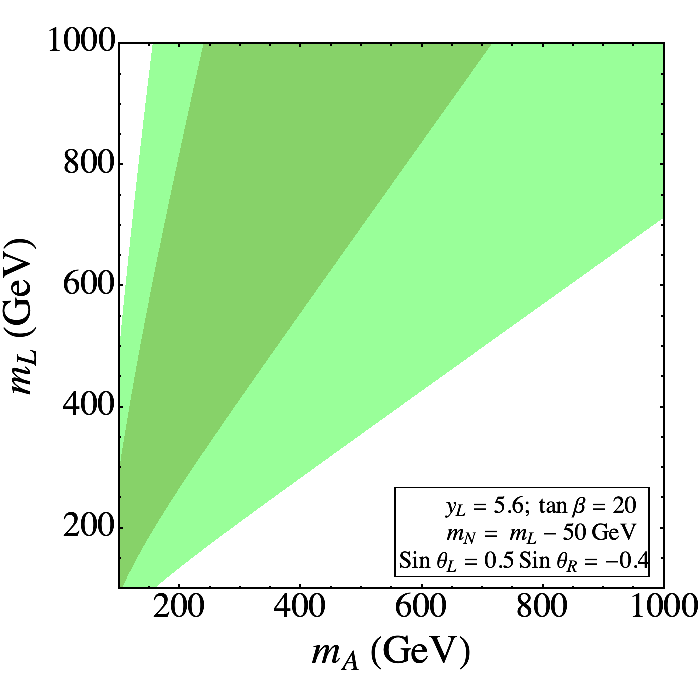} &
		\includegraphics[width=80mm, height=80mm]{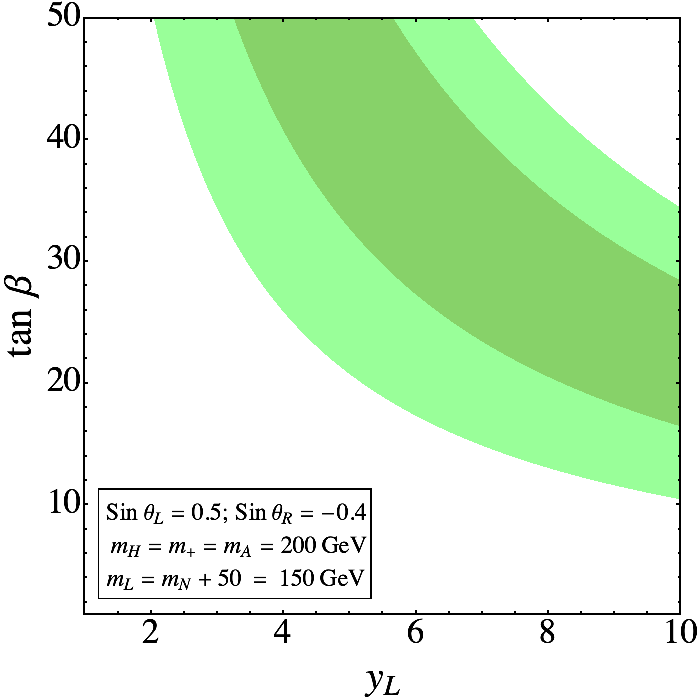}  \\
		(a) & (b) \\
		\includegraphics[width=80mm, height=80mm]{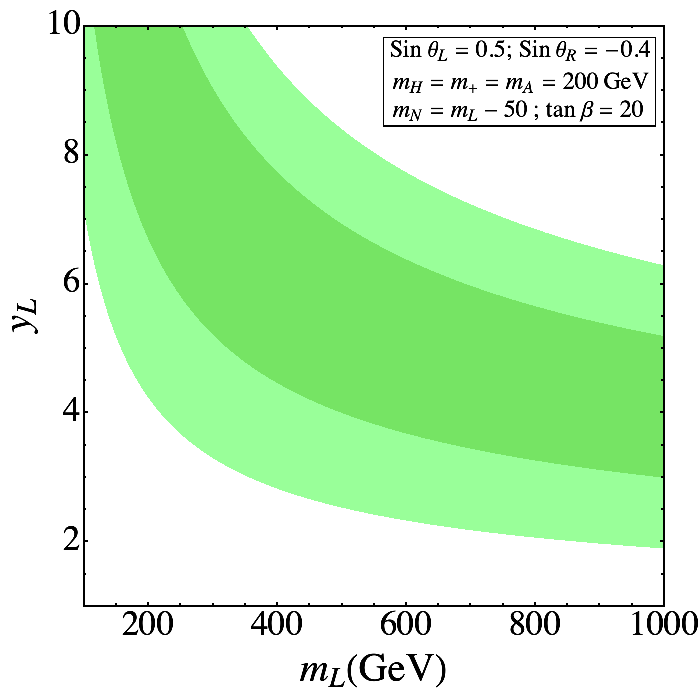} &
		\includegraphics[width=80mm, height=80mm]{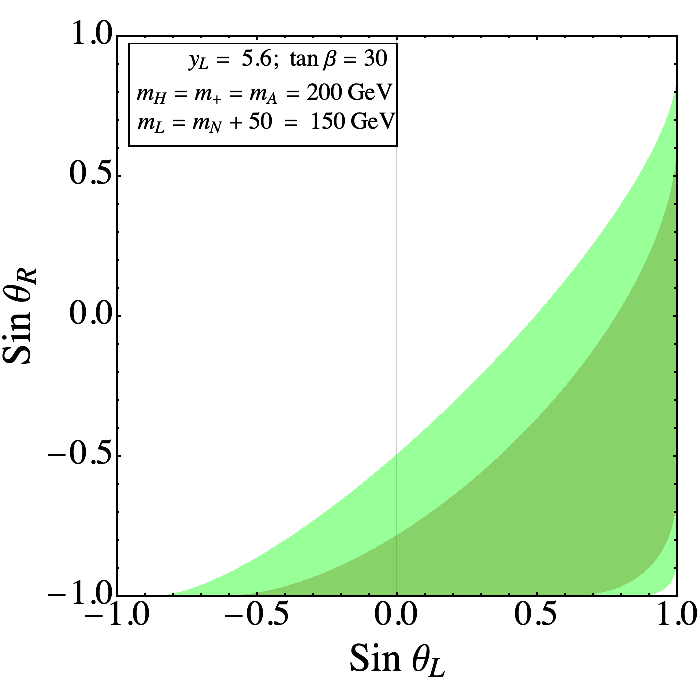} \\
		(c) & (d)
	\end{tabular}
	\caption{ Model parameter space satisfying the muon anomalous magnetic moment in Type-X 2HDM with VLLs at $1\sigma$ (dark green) and $2\sigma$ (light green) uncertainty. }
	\label{fig:muMM2}
\end{figure}
To further analyze the restrictions, we continue our investigation of the allowed space satisfying \gmin2 constraints in Fig. \ref{fig:muMM2}. We show the allowed
parameter space in (a) $(m_L-m_A)$, (b) $(y_L-\tan\beta)$ (upper panels) and (c) $(m_L-y_L)$, (d) $(\sin \theta_L-\sin \theta_R)$  (lower panels)  planes, respectively, for the Type-X + VLLs scenario only. Similar to Fig.~\ref{fig:muMM}, the dark and light shades of green 
reflect the $1\sigma$ and $2\sigma$ restrictions on the respective parameter space. In all the plots, we have considered equal masses for the heavy  and  charged Higgs bosons as $(m_A=m_H=m_+)$, and the vector-like neutrino mass $m_N$ lighter than that of the charged VLLs by 50 GeV. The Yukawa coupling $y_L$ is chosen to obey the perturbativity constraint $y_L < 4 \pi$.  As expected, the VLL Yukawa coupling $y_L$ is required to be large at smaller $\tan\beta$ (see Fig.~\ref{fig:muMM2}(b)) to yield a significant contribution from VLL Barr Zee loop, while for fixed $\tan\beta$, smaller $y_L$ can accommodate large mass for the VLLs $m_L$ (see Fig.~\ref{fig:muMM2}(c)) since both the Barr Zee diagrams in Fig.~\ref{fig:BarrZee}(a,d) are proportional to the product $\tan\beta\, y_L \, m_L^2$. One important observation is that the VLL mixing angles $\sin \theta_L$ and $ \sin \theta_R$ are required to have opposite signs,  with the right-handed mixing angle $\theta_R$ being negative. The alternate sign  choice, i.e. choosing $\theta_L$ to be negative will result in negative contribution to the \gmin2.  This is predominantly due to the relative negative sign in the VLL Yukawa couplings of the CP-odd Higgs with respect to 
 that of CP-even Higgs, as shown in Eq.~(\ref{eq:VLLparam}).
This justifies our choice of mixing angle in this study that also satisfies other constraints. Finally, our results in Fig.~\ref{fig:muMM2} illustrate the features of the model parameters that explain the \gmin2 anomaly.

\begin{figure}[htbp!]
	\centering
	\includegraphics[width=80mm, height=80mm]{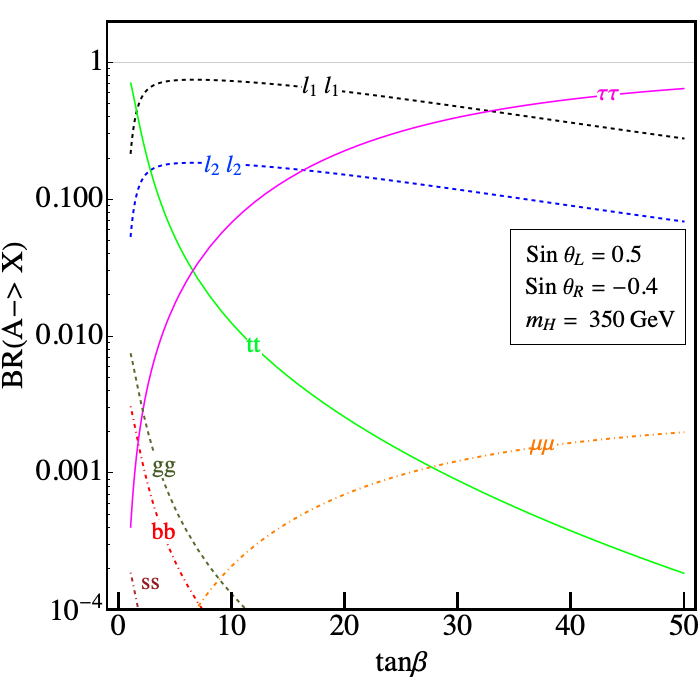} \hfill
	\includegraphics[width=80mm, height=80mm]{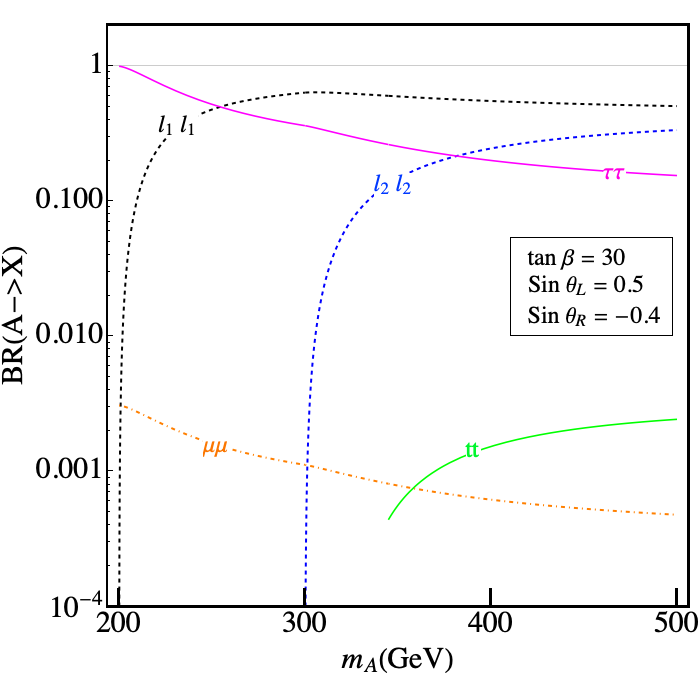} \\
	\includegraphics[width=80mm, height=80mm]{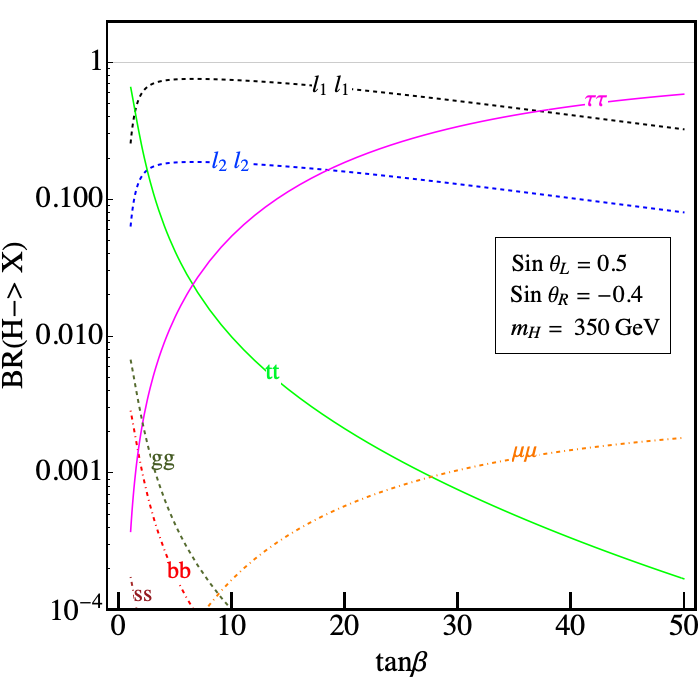} \hfill
	\includegraphics[width=80mm, height=80mm]{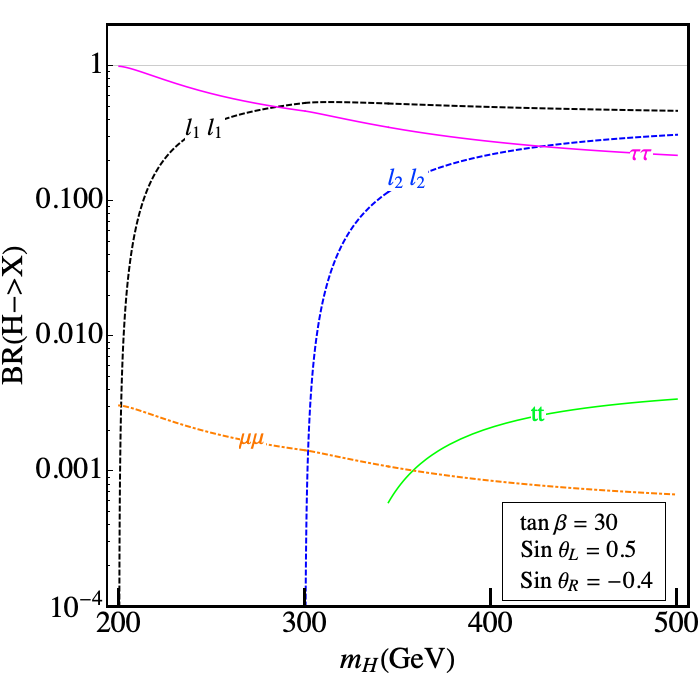}
	\caption{ Branching ratio of non-SM CP-odd (top) and CP-even Higgs bosons (bottom)  as functions of $\tan \beta$ for a fixed mass $m_H=m_A=350$ GeV (left panels) and as functions of mass, for fixed $\tan\beta=30$ (right panels). }
\label{fig:BRHA}
\end{figure}

Given our assumptions on non-SM mass limits,  we  address the latest constraints from the heavy Higgs boson searches at the  LHC Run II~\cite{Aad:2020zxo}. In the large $\tan\beta$ limit,
the heavy Higgs bosons in Type-X 2HDM become leptophilic and decay to $\tau\tau$ mode with almost 100\% branching ratio (BR). After the 
second run of LHC at 13 TeV center of mass energy with $139~\rm fb^{-1}$ luminosity, the ATLAS collaboration set new strong limits on such models from the heavy Higgs searches into $\tau\tau$ mode~\cite{Aad:2020zxo}, ruling out  heavy Higgs masses 
less than 1 TeV when  $\tan\beta \sim 20$. This imposes a stringent limit on the Type-X model parameter region consistent with \gmin2. However, consistent with our model assumptions, the heavy Higgs bosons can  decay, predominantly or significantly, into the vector-like charged leptons reducing the BR to $\tau \tau$ mode considerably. In Fig.~\ref{fig:BRHA}, we
show the BR of the CP-even heavy Higgs with mass $m_H$ (bottom panels) and CP-odd Higgs with mass $m_A$ (top panels) in the mass range where the decay into VLLs  is kinematically possible, for fixed mass $m_H=m_A=350$ GeV (left panels) and fixed $\tan\beta=30$ (right panels). 
As one can see, the VLL modes, if not dominate, then at least reduce the branching ratio to the $\tau \tau$ mode considerably, relaxing the LHC constraint, and ensuring the viability of the parameter region satisfying \gmin2 constraints.

Since, we have not considered any mixing between the VLLs and the SM leptons, the only possible decay modes for the VLLs are via gauge interaction
$(\ell_i^+ \to W^+ + N)$ yielding three body decays. The neutral lepton $N$ will contribute to the missing energy signal. A somewhat related 
study has been done in a previous work~\cite{Bahrami:2016has} where the possibility of detecting such VLL in the context of left-right symmetric model was explored. Moreover, in that paper, the possibility of the neutral heavy lepton being a viable Dark Matter candidate has also been discussed. Here, we leave such detailed study of LHC constraint in the 2HDM with VLLs for future exploration. 

\clearpage

\section{Summary and Conclusion}
\label{sec:summary}

We have shown that enlarging the fermionic content of the SM to include one generation of vector-like leptons (two doublets,  and two singlet charged leptons, one with SM quantum numbers, and the second a mirror, with opposite chirality) can resolve the 
 inconsistency between the theoretically expected and experimentally measured  values of the muon magnetic moment. We choose to do this in a scenario with a minimum number of parameters. As we do not allow mixing between vector-like and SM fermions (which would be restricted to involve third generation only, to avoid leptonic FCNCs) the simplest scenario that will obey all experimental constraints would be an extension of the SM by an additional scalar doublet, the 2HDM.
To reconcile the discrepancy, we concentrated on the soft $Z_2$-symmetry breaking 2HDM scenario 
 in Type-X (lepton specific) variant of the model, with explicit CP conservation in the potential, while also providing an analysis for Type-II scenarios. We restrict ourselves to working in the alignment limit, where the lightest CP-even Higgs boson coincides with that of the SM, insuring agreement with LHC data.
 
We first ensured that we worked in a parameter region (scalar and VLL masses, $\tan \beta$, and left and right VLL mixing angles) which satisfies constraints from precision electroweak parameters, $S$ and $T$, SM Higgs data (in particular, we guarantee the agreement with Higgs decay to diphotons, which is affected by the additional scalars and VLLs), while requiring the coupling constants to be within limits respecting the  perturbativity, unitarity and vacuum stability of the Higgs potential.
 
 Within these limits, we then investigated the effects of the additional scalars and VLLs on the anomalous magnetic moment of the muon.
 At one-loop level, in addition to the SM contributions, the additional Higgs bosons also enter in the loops, while  (in addition to the non-SM scalars of the 2HDM potential),  the VLLs contribute significantly to the \gmin2 only at  the two-loops, via the Barr Zee diagrams. 
 
 We performed a comprehensive analysis, including 
 all  relevant one- and two-loop diagrams, and showed that the allowed parameter space for  Type-X 2HDM  is greatly enhanced 
from  its  version without VLLs. Previously, it was shown that in Type-X 2HDM  without VLLs only very light pseudoscalar masses around $m_A < 60$ GeV, valid for $\tan\beta \sim 100$, can be consistent with the \gmin2, while the addition of VLLs allows the mass of the pseudoscalar Higgs to be as large as ${\cal O} (1~\rm TeV)$ with even  $\tan\beta \sim 50$, and lighter masses release more parameter space for $\tan \beta$. Moreover, introduction of VLL allows some parameter space for the Type-II 2HDM. Type-X and Type-II share the same lepton Yukawa coupling, but in Type-II lepton and down quark Yukawa couplings are correlated.  When these are large, flavor bounds from the quark sector restrict further the parameter space for Type-II 2HDM. Without VLLs, none survives, while introducing VLLs opens the same parameter regions as for Type-X, before additional restrictions apply. 

We also discussed the recent restrictions on the heavy Higgs masses from the LHC. There, we find that the
 additional decay modes of the heavy Higgs into the vector-like leptons, allowed in our parameter space,  relax the recent strong bound from Higgs decay into ditaus on the Type-X and Type-II 2HDM, thus not diminishing the parameter region consistent with \gmin2 agreement. 
 
As a passing note, we would like to comment that the potential of observing VLLs at colliders is outside the scope of the present paper but has been investigated more thoroughly in other works~\cite{Capdevilla:2020qel}. In our model, as the VLLs do not mix with third generation leptons,  they can be pair-produced at the LHC and decay via a virtual $W$ boson into a VLN (seen as missing energy), one neutrino and one ordinary lepton, yielding a $2 \ell+E_{\rm miss}$ signal.

\section*{Acknowledgements}
The work of M.F. has been partly supported by NSERC through grant number SAP105354. The work of I.S. was supported by World Premier International Research Center Initiative (WPI Initiative), MEXT, Japan.
\bibliographystyle{JHEP}
\bibliography{ref.bib}
\end{document}